\input{aipcheck}

\documentclass[natbib,final]{aipproc}

\layoutstyle{6x9}

\begin{document}

\title{Gamma-ray burst observations with new generation imaging atmospheric Cerenkov Telescopes in the FERMI era}


\author{S. Covino}{
  address={INAF / Brera Astronomical Observatory, Via Bianchi 46, 23807, Merate (LC), Italy}
}

\author{M. Garczarczyk}{
  address={Max-Planck-Institut f\"ur Physik, F\"ohringer Ring 6, D-80805 M\"unchen, Germany}
  ,altaddress={IFAE, Edifici Cn., Campus UAB, 08193 Bellaterra, Spain} 
}

\author{N. Galante}{
  address={Max-Planck-Institut f\"ur Physik, F\"ohringer Ring 6, D-80805 M\"unchen, Germany}
}

\author{M. Gaug}{
  address={Instituto de Astrofisica de Canarias, via L\'actea s/n, 38205 La Laguna, Tenerife, Spain}
}

\author{A. Antonelli}{
  address={INAF / Rome Astronomical Observatory, Via Frascati 33, 00044, Monte Porzio (Roma), Italy}
}
  
\author{D. Bastieri}{
  address={Universit\`a di Padova and Istituto Nazionale di Fisica Nucleare (INFN), 35131, Padova, Italy}
}
    
\author{S. Campana}{
  address={INAF / Brera Astronomical Observatory, Via Bianchi 46, 23807, Merate (LC), Italy}
}

\author{F. Longo}{
  address={Dipartimento Fisica and INFN Trieste, 34127 Trieste, Italy}
}

\author{V. Scapin}{
  address={Universit\`a di Udine, and INFN Trieste, 33100 Udine, Italy}
}

\begin{abstract}
After the launch and successful beginning of operations of the FERMI satellite, the topics related to high-energy observations of gamma-ray bursts have obtained a considerable attention by the scientific community. Undoubtedly, the diagnostic power of high-energy observations in constraining the emission processes and the physical conditions of gamma-ray burst is relevant. We briefly discuss how gamma-ray burst observations with ground-based imaging array Cerenkov telescopes, in the GeV-TeV range, can compete and cooperate with FERMI observations, in the MeV-GeV range, to allow researchers to obtain a more detailed and complete picture of the prompt and afterglow phases of gamma-ray bursts.
\end{abstract}

\maketitle

\section{Introduction}

High Energy (HE) or Very High Energy (VHE) observations in the MeV-GeV-TeV range of Gamma-Ray Bursts (GRBs) have been suggested to be powerful diagnostic for the emission processes and physical conditions of GRBs by many authors \cite{FaPi08,Fal08,Ahar08,DeFr08,GaPi08,LeDe08}. As a matter of fact, in spite of the many successes in the interpretation of the various phases of a GRB \cite{ZhMe03,Zh07}, there are still fundamental unanswered questions involving essentially all aspects of the GRB phenomenology. 

In the past, several detections of GRBs in the MeV-GeV range were performed by the Energetic Gamma-Ray Experiment Telescope (EGRET) aboard  the Compton Gamma-Ray Observatory (CGRO)\footnote{http://heasarc.gsfc.nasa.gov/docs/cgro/}. Five of these events were detected at energies above 30\,MeV \cite{Din95} and a total of about 50 photons were collected. HE emission was observed roughly coincident in time with the prompt emission (i.e. for GRB\,940217 \cite{Hurl94}) or delayed and more naturally associated to the afterglow (i.e. GRB\,940217 and GRB\,941017 \cite{Hurl94, Gonz03}). Indeed, these HE observations are likely the first real discovery of a GRB afterglow, in advance of the \textit{Beppo}SAX\footnote{http://www.asdc.asi.it/bepposax/} discovery of the soft X-ray afterglow in 1997 \cite{Cos97}. These sparse observations already provided some important observational facts. In the only case when emission simultaneous to the prompt phase was detected (GRB\,040217) the HE count rate exceeded the count rate during the later afterglow. However, the HE emission, including a single 18\,GeV photon received almost two hours after the GRB prompt, had a markedly different time evolution compared to the lower-energy afterglow, with a longer duration and a rather stable count-rate. In the case of GRB\,941017, again the HE component showed a slower evolution compared to the low-energy afterglow and moreover carried at least three times more energy than the low-energy afterglow.

More recently, MeV-GeV detections of GRBs have been also obtained by AGILE\footnote{http://agile.rm.iasf.cnr.it/} \cite{Giul08} and FERMI\footnote{http://fermi.gsfc.nasa.gov/}.

On the contrary, in spite of continuous attempts, so far no convincing detections at higher energies ($\sim$ TeV) have been obtained. The Milagrito experiment claimed to have observed emission at about 0.1\,TeV for GRB\,970417A \cite{Atk03} at a $\sim 3\sigma$ significance, however the Milagrito successor, Milagro\footnote{http://www.lanl.gov/milagro/index.shtml}, did not detect any significant signal in more than 50 GRBs observed. 

\begin{figure}
\centering\includegraphics[width=\textwidth]{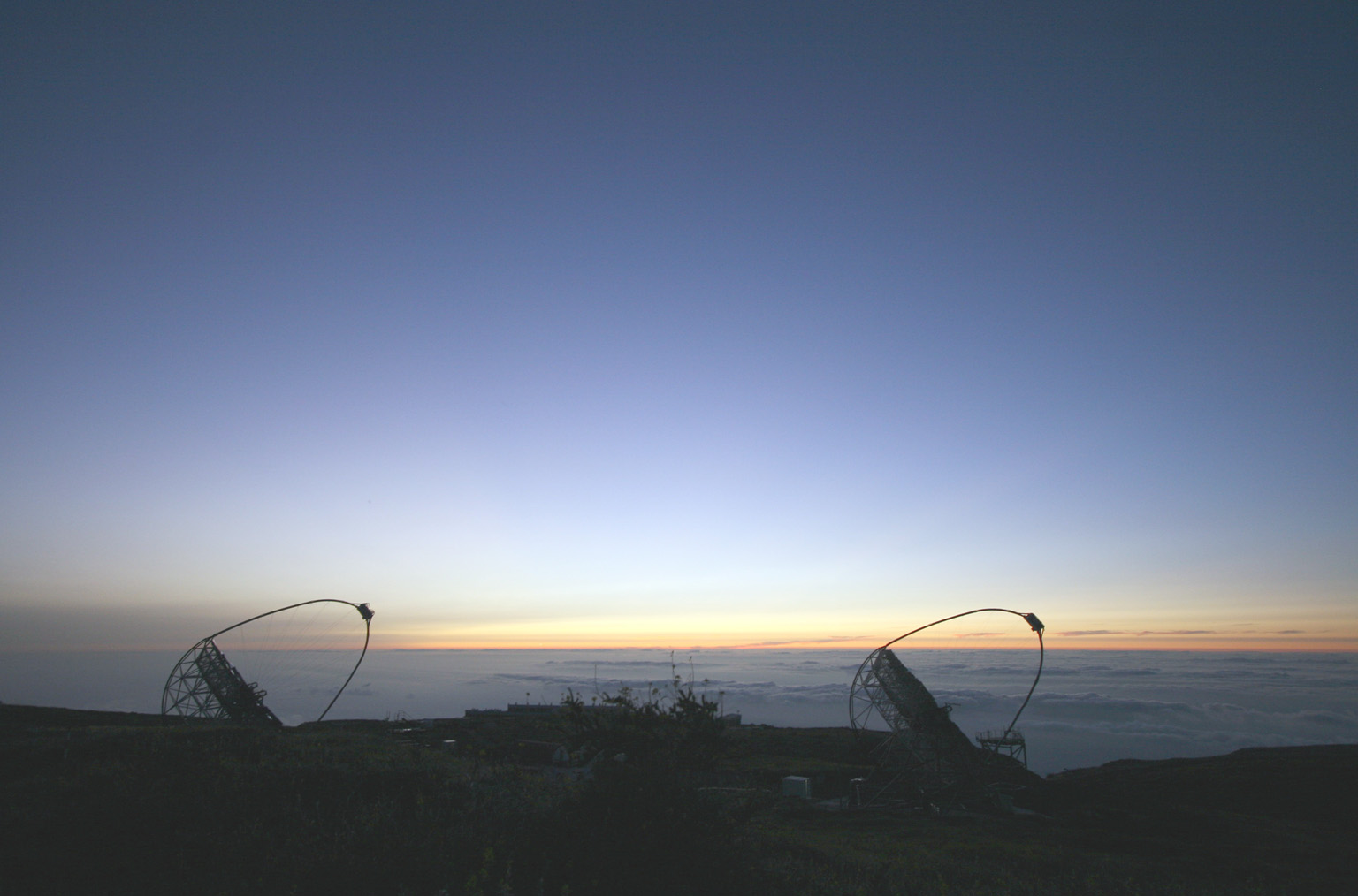}
\caption{A pictorial view of the MAGIC and MAGIC-II telescopes at Roque de los Muchachos, Canary Islands (from the MAGIC web site).}
\label{fig:magic}
\end{figure}

Null detections so fare have also been reported by various Imaging Atmospheric Cerenkov Telescopes (IACTs) as HESS\footnote{http://www.mpi-hd.mpg.de/hfm/HESS/}, VERITAS\footnote{http://veritas.sao.arizona.edu/} and in particular MAGIC\footnote{http://wwwmagic.mppmu.mpg.de/index.en.html} (Fig.\,\ref{fig:magic}) which is characterized by the lowest energy threshold \cite{Alb06,Mor07,Bast07,Gal08,Gar08} which has been brought down to 25\,GeV under favourable conditions in a recent upgrade\cite{Aliu08}.

\section{Universe transparency at VHE}

The universe at VHE is not transparent. Photon with energies above above tens of GeV can interact with the diffuse infrared photon background, or Extragalactic Background Light (EBL) \cite{Hau01}. The optical depth $\tau$ for a $\gamma \sim 100$\,GeV photon at $z \sim 1$ can be as high as $\tau \sim 6$, depending on the specific model prediction. Observationally, for redshifts higher than about 0.2, the attenuation due to EBL interaction is uncertain, and the recent detection by the MAGIC telescope of the quasar 3C\,279 ($z=0.54$) at energies above 300\,GeV \cite{Alb08} could support more optimistic predictions with optical depth at $z\sim1$ not far from unity. 

\begin{figure}
\centering\includegraphics{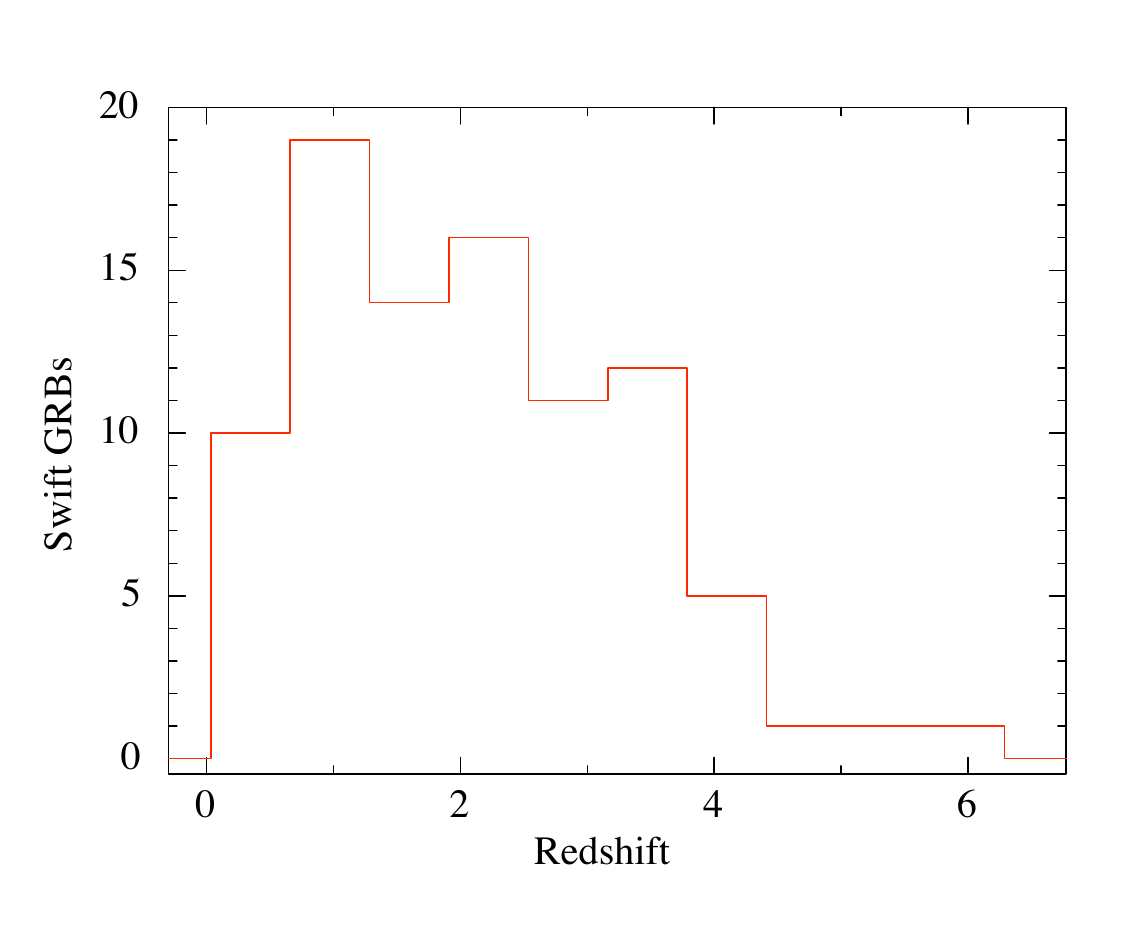}
\caption{Histograms of the GRBs localized by \textit{Swift} till Fall 2008 and with a measured redshift. The median value is about 2.3 and, in any case, the percentage of GRBs at redshift lower than $\sim  1$ is small, although clearly not negligible.}
\label{fig:swiftz}
\end{figure}

Most of the GRBs observed by IACTs are follow-up observations of alerts distributed by the \textit{Swift} satellite\cite{Geh04} over the GRB Coordinates Network\footnote{http://gcn.gsfc.nasa.gov/} (GCN). Histogram in Fig.\,\ref{fig:swiftz} shows the redshift distribution of the GRBs detected by \textit{Swift} till Fall 2008. The majority of \textit{Swift} GRBs are at redshift substantially larger than one. The conclusion is that almost independently of the predicted flux at VHE from a GRB, unless the event is relatively nearby, the EBL attenuation prevents almost any hope to detect emission. Given, however, the diagnostic importance of these observations this also suggests to devote any effort to try to catch all promptly visible GRBs with high priority compared to other targets in order not too loose those few at relatively low redshift.

FERMI is also delivering well localized GRB alerts, although at a lower rate compared to \textit{Swift} ($\sim 20$/year). It is possible that the relatively hard energy range covered by FERMI makes it more sensitive to nearby objects increasing the probability of observing low-redshift GRBs.

\section{GRB HE emission processes}

Many good reviews about GRB interpretative scenarios are available. The leading picture is the so-called cosmological fireball model \cite{Pir99,ZhMe03,Mes06,Zh07}. There are alternative scenarios presenting various issues of interest (the so-called "cannonball'' scenario \cite{Dar06}, the "fireshell'' scenario \cite{BiRu08}, etc.). However, for simplicity, and with no intention to be exhaustive, we briefly follow the main predictions for the fireball model, which is also the best studied and for which most of theoretical considerations are available.

In the fireball scenario the various observed phenomenologies from a GRB are essentially due to an ultrarelativistic outflow (initial Lorentz $\Gamma$ factor between 100-1000 \cite{Mol07}) generated during the final collapse of a high mass star or the coalescence of a binary system made by compact objects. The former case is usually associated to the long duration GRBs, while the latter to the short duration class. Independently of details, the emission is supposed to happen in two distinct regions: (1) relatively nearby the central engine ($\sim 10^{13}$\,cm) due to inhomogeneities in the outflow - the internal shocks, and (2) much further ($\sim 10^{16}$\,cm), when the outflow interacts with the circumburst medium - the external shocks. The theory for ultrarelativistic shocks is far from being well developed, however, it is assumed that electrons (and other charged particles) are accelerated in the shocks by first- and second-order Fermi processes and the electrons cool emitting synchrotron radiation. Coupling the dynamical evolution of the outflow with the micro-physics parameters driving the generation of synchrotron photons, light-curves and time-resolved spectra for the whole evolution of a GRB can be obtained. A comprehensive discussion about the strong and weak points of this model is clearly beyond the purpose of this proceeding, however this scenario has the merit to have provided a satisfactory general description of the observations although an increasing degree of complexity seems to be required to cope with the best quality available datasets.

Synchrotron spectra intrinsically cover a wide energy range, and for GRBs the high energy cutoff is mainly dependent on the Lorentz $\Gamma$ factor and it is derived comparing the electron cooling time to the acceleration time:

\begin{equation}
hv_{\rm MAX} \approx 30\Gamma/1+z \quad {\rm MeV}.
\end{equation}

With typical parameters of the prompt phase we could have photons up to a few GeV while in the early afterglow the photons extend up to energies of 10-100\,MeV. Therefore already synchrotron emission can produce photons in the HE range as observed during the prompt phase in the past. On the contrary, GeV photons in the afterglow require some additional mechanism. The observation of the high-energy cutoff of the synchrotron component therefore could allow researchers to estimate the bulk fireball Lorentz factor $\Gamma$ and, in turn, the observation of a cutoff at much lower energies would instead strongly question the interpretation of the GRB radiation as due to the synchrotron process.

Beyond electron synchrotron emission at the shock front of a GRB, protons are also accelerated and can 
cool via synchrotron emission. The expected energy range can be in TeV regime, but it is not clear how important proton synchrotron could be compared to other emission processes in this range\cite{Tot98,FaPi08,Asa08}.

The most important HE emission process for GRBs is likely Inverse Compton (IC) which is expected to be important in a large fraction of the micro-physical parameter space generating a synchrotron component \cite{PeWa04,PeWa05}. Depending on the specific conditions, it is possible to have multiple order IC scattering and at VHE energies the Klien-Nishina cross-section suppression has to be taken into account \cite{RiLi79}.

The cosmological fireball scenario offers a rich set of possibilities for an effective IC process. Synchrotron photons can be upscattered by the same electrons which cooled by synchrotron emission (Synchrotron Self-Compton, SSC). Electron and photon fields are isotropic and it is expected a temporal correlation between the synchrotron and IC emission. For typical parameters first and second order scatterings are often still in the Thomson regime \cite{FaPi08}. SSC can occur both during the internal and the external shocks effectively producing a HE component following the time evolution of the underlying synchrotron component. The importance of the IC is usually parametrized with $Y_{\rm IC} \approx U'_\gamma/U'_{\rm B}$, the comoving photon to magnetic field energy density ratio. The typical SSC photon energy during the internal shocks is, following \cite{FaPi08}:

\begin{equation}
hv_{\rm SSC} \approx 240\,{\rm GeV}~(1+Y_{\rm SSC})^{1/2}~R_{13}~(\epsilon_{\rm peak}/100\,{\rm KeV})^2  / L_{\rm syn,50},
\end{equation}

where $R$ is the distance from the central engine where the internal shocks occur, $\epsilon_{\rm p}$ is the synchrotron spectrum peak energy and $L$ the luminosity of the synchrotron component. However, due to pair production in the emitting region, a cutoff energy for the escape of IC photons can be derived:

\begin{equation}
hv_{\rm cut} \approx 2\,{\rm GeV}~(\epsilon_{\rm peak}/100\,{\rm KeV})^{(2-p)/p}~\delta t_{\rm v}^{2/p}~\Gamma_{0,2.5}^{(2p+8)/p},
\end{equation}

where $p$ is the power-law electron distribution index ($N(E) \propto E^{-p}$, typically $p \sim 2.3$), $\delta t_{\rm v}$ is the typical variability time and $\Gamma_0$ is the initial Lorentz factor. Essentially we can derive two important pieces of information. First of all, for a reliable prediction of a SSC component, we need to know the main parameters of the underlying synchrotron component, therefore requiring a real multiwavelength approach. Independently of the many possible details, the internal shocks (i.e. the prompt phase) is not, in general, the best place where to look for VHE emission due to the intense internal opacity preventing VHE photons to freely escape. A possible way to overcome this limitation is for GRBs characterized by a very high bulk Lorentz factor $\Gamma > 10^3$. In this case the internal shock radius is substantially farther from the central engine, the variability time is proportionally longer and the larger volume available in the emitting region reduces the opacity. We mention in passing that one of the possible interpretations of the multiwavelength observations of the recent exceptionally bright GRB\,080319B \cite{KuPa08,Rac08,Pan08a} is that the optical emission is due to synchrotron and the keV/MeV emission is due to first order SSC. For typical parameters, in this case, the second order IC component could have carried 10-100 more energy than the lower energy components and could have been easily detectable by the AGILE satellite or by ground-based IACTs as MAGIC. Very unfortunately, the field of view of GRB\,080319B was occulted by the Earth from the AGILE location and the \textit{Swift} alert came at the MAGIC site (Canary Islands) when the morning twilight already begun. 

It might also happen that electrons accelerated at the shock front upscatter photons coming from different regions (External Inverse Compton, EIC), allowing many different possibilities (i.e. photons from reverse shock or delayed internal shocks interacting with electrons in the forward shock, etc. ). In this case the temporal relation between the synchrotron and the IC component gets more complicated since now in the comoving frame photons are strongly beamed while electrons are still isotropic and the angular dependence of the IC scattering \cite{RiLi79} can substantially delay (and lower, compared to the isotropic case) the total IC flux. This scenario is particularly intriguing for X-ray flares, which are frequently observed during the afterglow phase of some GRBs \cite{Chi07,Fal07}. The impulsive emission generated by the flare at low (keV/MeV) energy can be converted to a long-lasting component at VHE \cite{GaPi07,FaPi08,GaGue08}. Moreover, in case flares are due to interactions occurring farther from the central engine than the internal shocks, VHE photons can more freely escape due to the reduced opacity. In general the richest set of possibilities is during the early-afterglow, i.e. between 10-1000\,s from the burst, when both the forward and the short-duration reverse shocks coexist \cite{SaEs01,ZhMe01,GaPi08}. Both SSC and EIC are usually possible and in general predictions for VHE flux depends on the details of micro-physical parameters generating the synchrotron emission. Moreover, all the ingredients which often are added to the standard fireball model (late energy injection, evolving micro-physical parameters, circumburst medium density variation, etc.) to cope with the observations are potentially able to affect the predictions for VHE components, although IC emission in the GeV range should likely be a common feature.

Many other processes can be important for GRBs both during the prompt and the afterglow phases.
IC due to electrons moving relativistically into a soft photon background (bulk Compton scattering), as it would be the case for mass ejected relativistically crossing a pre-existent photon field, is also an often invoked emission process\cite{ShDa95,Laz04,Pan08b}. An interesting source of VHE photon can come from neutral pion decay and charged pion synchrotron \cite{Katz94,DeAt04}. In both cases the peak energy is in the TeV range, although it is not clear the relative importance of these components compared to the other possible emission processes.

In the cosmological fireball model the outflow is driven by the internal energy of the fireball. However, it is possible that the outflow is instead energetically dominated by large scale ordered magnetic fields, perhaps advected from the central source \cite{Usov94,ZhKo05,Lyu06}. In this case GRBs could be the first case of astrophysical objects, where the energy content of which is dominated by magnetic energy. These family of models are of relevant importance to overcome some of the difficulties of the standard model in particular during the prompt phase \cite{ZhKo05}. Regarding the VHE emission, in general a magnetically dominated outflow should produce a weaker IC component and therefore being less effective in producing VHE photons \cite{FaPi08}.

\section{FERMI and IACTs}

Although the perspectives for observations of VHE emissions from GRBs are promising, it is clear that most of the emission processes mentioned before are more effective in generating GeV photons than higher energy (TeV) emission. This is particularly important if we also consider the attenuation effect of the EBL interaction for all but the most nearby sources. The energy range covered by the instruments onboard FERMI give an effective area of $\sim 10^4$\,cm$^2$ at about 100\,GeV while for an IACT like MAGIC the effective area at about 100\,GeV is of the order of a few$\times 10^4$\,m$^2$. A comparison is indeed difficult because clearly the duty cycle of a satellite is much better than any ground based instrument, the performances of which are strongly dependent on the Lunar phase, weather, etc. Nevertheless, for almost any reasonable spectrum VHE photons are intrinsically rare compared to lower energy photons and the higher the energy is the larger the required effective area should be. For a moderately bright (fluence $\sim 10^{-5}$\,erg\,cm$^{-2}$ in the \textit{Swift}-BAT range) GRB at $z \sim 1$ it is easy to estimate the number of photons FERMI can collect at MeV or GeV energies due to the various discussed processes \cite{FaPi08,LeDe08}. In general, apart from very bright events (like GRB\,080319B, the probability of which was roughly estimated to $\sim$ one/10 years \cite{Rac08}), FERMI should easily detect a few tens of photons at energies in the MeV range, while just a few photons can be detected in GeV range. On the contrary, in the tens of GeV range an IACT can again detect tens of photons. It appears clear that apart from the mere detection, for a reliable spectral analysis with the perspective to single out which emission process is efficient, the simultaneous availability of FERMI and an IACT is mandatory. Not only the statistics of detected photons is substantially improved, but the energy range leverage makes any analysis much more constraining. A low energy cutoff as low as  50\,GeV seems to be a required feature to improve the capabilities of an IACT to detect emission from a GRB. The foreseen improvements of MAGIC, HESS and VERITAS seem all be able to guarantee an effective area of several tens of square meters at a few tens of GeV. Low energy threshold observations, although not for the typical GRB observing conditions (unpredictable zenith angle and atmospheric transparency, etc.), have already obtained by MAGIC \cite{Aliu08}. Morever, the rapid pointing capabilities of the new generation of IACTs, in particular of MAGIC which already now can point in a few tens of seconds everywhere in the visible sky, increases the chances to track a GRB field during the most promising phases for VHE emission, the end of the prompt phase and the early-afterglow.

\section{Conclusions}

In this proceeding we have briefly reminded the main emission processes proposed to be effective during the varios phases of a GRB. The discussion was developed with the cosmological fireball model as a reference, but some of the results can be applied in other alternative scenarios. 

The main conclusion is that the availability of an operational satellite, FERMI, devoted to HE observations offers the best opportunities for routine detections of HE photons from a GRB. However, for a full spectral analysis able to disentangle the huge varieties of processes which can be effective in producing VHE emission from a GRB, a real multiwavelength approach is mandatory. Low energy observations (radio, optical, soft X-rays) are required to define the synchrotron spectrum parameters which in turn can allow to better constrain the HE and VHE predictions. Then VHE observations are required as well to allow a spectral analysis of the HE components detected by FERMI and to allow time-resolved analysis in case the low-energy threshold is low enough.

For IACT managers, a GRB detected by the FERMI-LAT should be considered a high-priority target to be observed as long as FERMI is following it. HE and VHE emissions do not necessarily follow the same time evolution of the lower energy component. 

Optical telescopes, possibly robotic and with rapid pointing capabilities, should be considered standard facilities for new generation IACTs, providing observations (not only for GRBs) fundamental for a profitable interpretation of the HE and VHE components.

\end{document}